# Using an Instrumental Variable to Test for Unmeasured Confounding

Zijian Guo[a], Jing Cheng[b], Scott A. Lorch[c], Dylan S. Small[a]

An important concern in an observational study is whether or not there is unmeasured confounding, i.e., unmeasured ways in which the treatment and control groups differ before treatment that affect the outcome. We develop a test of whether there is unmeasured confounding when an instrumental variable (IV) is available. An IV is a variable that is independent of the unmeasured confounding and encourages a subject to take one treatment level vs. another, while having no effect on the outcome beyond its encouragement of a certain treatment level. We show what types of unmeasured confounding can be tested for with an IV and develop a test for this type of unmeasured confounding that has correct type I error rate. We show that the widely used Durbin-Wu-Hausman (DWH) test can have inflated type I error rates when there is treatment effect heterogeneity. Additionally, we show that our test provides more insight into the nature of the unmeasured confounding than the DWH test. We apply our test to an observational study of the effect of a premature infant being delivered in a high-level neonatal intensive care unit (one with mechanical assisted ventilation and high volume) vs. a lower level unit, using the excess travel time a mother lives from the nearest high-level unit to the nearest lower-level unit as an IV. Copyright © 0000 John Wiley & Sons, Ltd.

## 1. Introduction

Randomized controlled studies are the gold standard to compare the effects of treatments between different treatment groups. Unfortunately, randomized controlled studies are often not feasible because of cost or ethical constraints. When randomized studies are not feasible, observational studies provide important evidence about the comparative effectiveness of different treatments. Since treatments were not randomly assigned, a major concern in an observational study is confounding, meaning that the treatment groups may differ before treatment in ways that also affect the outcome. If the confounders are measured, these differences can be adjusted for by matching, stratification or regression[1]. However, there is often concern that there are unmeasured ways in which the treatment groups differ that affect the outcome, meaning that there is unmeasured confounding. Even when there is unmeasured confounding, it is possible to obtain a consistent estimate of the causal effect of treatment for a certain sub-population (the compliers) if an instrumental variable (IV) can be found. An IV is a variable that (1) is independent of the unmeasured confounding; (2) encourages, but does not force, a subject to take one treatment level vs. another; and (3) has no effect on the outcome beyond its encouragement of a certain treatment level. For discussions of IVs, see [2, 3, 4, 5, 6, 7, 8, 9]. In this paper, we develop a method for using an IV to test whether there is unmeasured confounding. Detecting whether there is unmeasured confounding is valuable in many studies because if unmeasured confounding is found in a given study, it suggests that for studying related questions, researchers should either try to measure more confounders or seek to find IVs.

[a] *Department of Statistics, The Wharton School, University of Pennsylvania*
[b] *Division of Oral Epidemiology and Dental Public Health, School of Dentistry, University of California, San Francisco (UCSF)*
[c] *Department of Pediatrics, University of Pennsylvania*
\* *Correspondence to: Dylan Small, Department of Statistics, The Wharton School, University of Pennsylvania, 400 Huntsman Hall, Philadelphia, PA 19104, e-mail: dsmall@wharton.upenn.edu*

The existing and widely used test for whether there is unmeasured confounding using an IV is the Durbin-Wu-Hausman endogeneity test, hereafter called the DWH test, independently proposed by Dur bin [10], Wu [11] and Hausman[12]. The DWH test compares an estimate of the average treatment effect under the assumption that there is no unmeasured confounding to an estimate of the average treatment effect using an IV that allows for unmeasured confounding. The IV estimate of the average treatment effect is assumed to be consistent so that a significant difference between it and the estimate that assumes no unmeasured confounding is taken as evidence of unmeasured confounding. The two estimates of the average treatment effect in the DWH test assume that the treatment effect is homogeneous, meaning that the treatment effect is not different by covariates; see [13, 4, 14, 15, 16] for discussion of homogeneity assumptions. [15] noted that if the DWH test rejects, one cannot be sure whether it is because of unmeasured confounding or treatment effect heterogeneity. Extensions to the conventional DWH test have been developed that allow for heteroskedasticity[17, 18], but these tests all have the property that if they reject, one cannot sure be sure whether it is because of unmeasured confounding or treatment effect heterogeneity.

In this paper, we develop a test that distinguishes between unmeasured confounding and treatment effect heterogeneity. We discuss what types of unmeasured confounding can be tested for and provide a test with correct type I error rate for the testable types of unmeasured confounding. In addition to having the advantage over the DWH test of having correct type I error rate for testing unmeasured confounding when there is treatment effect heterogeneity, our testing approach also provides more insight into the nature of the unmeasured confounding by providing separate tests for two different types of unmeasured confounding. In the DWH test, these two types of unmeasured confounding are lumped together.

The motivating application for our work is an observational study of neonatal care that seeks to estimate the effect on mortality of a premature infant being delivered in a high-level neonatal intensive care unit (NICU) vs. a lower-level NICU. A high-level NICU is defined as a NICU that has the capacity for sustained mechanical assisted ventilation and delivers at least 50 premature infants per year. Estimating the effect of being delivered at a high-level NICU is important for determining the value of a policy of regionalization of perinatal care that aims for premature infants to be mostly delivered in high-level NICUs [19]. Regionalization of perinatal care was developed in the 1970s along with the expansion of neonatal technologies, but by the 1990s, regionalization began to weaken in many areas of the United States [20]. The difficulty in studying the causal effect of a premature infant being delivered in a high-level vs. a lower-level NICU is confounding by indication – the mother and the mother's physician may try harder to deliver an infant at a high level NICU if the infant is at higher risk of mortality. In our data from Pennsylvania (described in Section 7), the unadjusted death rate in high-level NICUs is *higher* than in low-level NICUs, 23 vs. 12 deaths per 1000 deliveries. Since it is unlikely that high level NICUs are worse than low level NICUs, the higher unadjusted death rate in high level NICUs probably reflects confounding by indication. Our data contains a number of potential confounders, including birth weight, month prenatal care started, mother's education and whether the mother went into labor prematurely. After adjustment for measured confounders by propensity score matching, the death rate is 0.5 deaths per 1000 deliveries lower in high-level NICUs [21]. However, we are concerned about unmeasured potential confounders, such as the severity of a mother's comorbid condition or an infant's antenatal condition, lab results, fetal heart tracing results, the compliance of the mother to medical treatment and the physician's history with the mother. These variables are known to the physicians who assess a mother's probability of delivering a high-risk infant. Based on this probability, the physicians then play a role in deciding where the mother should deliver. To attempt to deal with the problem of potential unmeasured confounding, we have collected data on a proposed IV, the excess travel time that a mother lives from the nearest high-level NICU compared to the nearest lower-level NICU; specifically, the IV is whether or not the mother's excess travel time is less than or equal to 10 minutes. Excess travel time to a hospital delivering speciality care has been used as an IV in other medical settings, such as studies of the effect of cardiac catheterization on survival in patients who suffered an acute myocardial infarction[22]. In obstetric care, prior work suggests that women tend to deliver at the closest hospital so that we expect that excess travel time will have a strong effect on where the infant is delivered[23]. We discuss in Section 7 reasons for thinking excess travel time is a valid IV. Our goal in this paper is to use the putative IV excess travel time to test whether there is unmeasured confounding in the study of the effect of high-level vs. lower-level NICUs. If unmeasured confounding is found, it suggests that previous studies of the effect of high-level vs. lower-level NICUs that assumed no unmeasured confounding provided biased estimates and that future studies of the effect of NICU level on mortality and related medical questions should seek to measure more confounders and/or find and measure IVs. Related medical questions include the effects of high level NICUs on complications, length of stay and readmissions, which have not yet been studied systematically.

The rest of the paper is organized as follows. In Section 2, we set up the causal framework and introduce notation and assumptions. In Section 3, we discuss what type of unmeasured confounding can be tested for when there are heterogeneous treatment effects. In Section 4, we develop our method for testing for unmeasured confounding using an IV. In Section 5, we discuss the DWH test and how it performs when there are heterogeneous treatment effects. In Section 6, we present a simulation study comparing our method to the DWH test. In Section 7, we apply our test to the study of the effect of high-level vs. lower-level NICUs. Finally, we provide conclusions in Section 8.

## 2. The Framework

### 2.1. Notation

The IV $Z$ and the treatment $A$ are assumed to each be binary, where level $0$ of the treatment is considered the "control" (lower-level NICU in our application) and level $1$ is considered to be the "treatment" (high-level NICU in our application). We let $\mathbf{Z}$ denote the $N$-dimensional vector of IV values for all $N$ subjects, with individual elements $Z_i = z \in \{0, 1\}$ for subject $i$; level $1$ of the IV is assumed to encourage receiving the treatment compared to level $0$. Let $\mathbf{A}^{\mathbf{z}}$ be the $N$-dimensional vector of potential treatment under IV assignment $\mathbf{z}$, with individual element $A_i^{\mathbf{z}} = a \in \{0, 1\}$ according to whether subject $i$ would take the control or treatment under $\mathbf{z}$. We let $\mathbf{Y}^{\mathbf{z}, \mathbf{a}}$ be the vector of potential responses that would be observed under IV levels $\mathbf{z}$ and treatment levels $\mathbf{a}$, with individual element $Y_i^{\mathbf{z}, \mathbf{a}}$ for subject $i$. $\{Y_i^{\mathbf{z}, \mathbf{a}}\}$ and $\{A_i^{\mathbf{z}}\}$ are "potential" responses and treatments in the sense that we can observe only one value in each set. We let $Y_i$ and $A_i$ be the corresponding observed outcome and treatment variables for subject $i$. We let $\mathbf{X}_i$ denote the measured confounders for subject $i$. We assume that $\mathbf{X}_i$ includes an intercept.

*2.2. Assumptions*

We make the same assumptions as [2] within strata of the measured confounders $\mathbf{X}$.

1. *Stable unit treatment value assumption (SUTVA)[29]: (a) If $\mathbf{z}_i = \mathbf{z}'_i$, then $A_i^{\mathbf{z}} = A_i^{\mathbf{z}'}$. (b) If $\mathbf{z}_i = \mathbf{z}'_i$ and $\mathbf{a}_i = \mathbf{a}'_i$, then $Y_i^{\mathbf{z},\mathbf{a}} = Y_i^{\mathbf{z}',\mathbf{a}'}$. This assumption allows us to write $Y_i^{\mathbf{z},\mathbf{a}}$ and $A_i^{\mathbf{z}}$ as $Y_i^{z,a}$ and $A_i^z$, respectively, for subject $i$.*
2. *IV is independent of unmeasured confounding: Conditional on $\mathbf{X}$, for a randomly chosen subject, the IV $Z$ is independent of the vector of potential responses and treatments $(Y^{0,0}, Y^{0,1}, Y^{1,0}, Y^{1,1}, A^0, A^1)$.*
3. *Exclusion restriction: For each subject $i$, $Y_i^{z,a} = Y_i^{z',a}$ for all $z, z'$ and $a$, i.e., the IV level affects outcomes only through its effect on treatment level. This assumption allows us to define $Y_i^a \equiv Y_i^{0,a} = Y_i^{1,a}$ for $a = 0, 1$.*
4. *IV affects treatment: $P(A = 1|Z = 1) > P(A = 1|Z = 0)$.*
5. *Monotonicity: $A_i^1 \geq A_i^0$ for all $i$. This assumption says that there is no one who would always do the opposite of what the IV encourages, i.e., no one who would not take the treatment if encouraged to do so by the IV level but would take the treatment if not encouraged by the IV level.*

We further make the consistency assumption[30]:

6. *Consistency: No matter how subject $i$ is administered the treatment, the potential outcome is the same so that:*

$$Y_i = A_i Y_i^1 + (1 - A_i) Y_i^0,$$

*where $Y_i$ and $A_i$ are the observed outcome and treatment for subject $i$.*

*2.3. Compliance Class*

Based on a subject's joint values of potential treatment $(A_i^0, A_i^1)$, a subject can be classified into one of four latent compliance classes [2]:

$$C_i = \begin{cases} nt \text{ (never-taker)} & \text{if } (A_i^0, A_i^1) = (0,0) \\ co \text{ (complier)} & \text{if } (A_i^0, A_i^1) = (0,1) \\ at \text{ (always-taker)} & \text{if } (A_i^0, A_i^1) = (1,1) \\ de \text{ (defier)} & \text{if } (A_i^0, A_i^1) = (1,0) \end{cases}$$

Under the monotonicity assumption, there are no defiers. We can observe only one of $A_i^0$ and $A_i^1$, so a subject's compliance class is not observed directly but it can be partially identified based on IV level and observed treatment as shown in Table 1. Based on Table 1, the following quantities are identified under assumptions $1 - 6$ based on the observable data: $P(C = at) = P(A = 1|Z = 0)$, $P(C = nt) = P(A = 0|Z = 1)$, $P(C = co) = 1 - P(A = 1|Z = 0) - P(A = 0|Z = 1)$, $E(Y^1|C = at) = E(Y|Z = 0, A = 1)$, $E(Y^0|C = nt) = E(Y|Z = 1, A = 0)$, $E(Y^1|C = co) = \frac{E(Y|Z=1,A=1) - [P(C=at)/\{P(C=at)+P(C=co)\}]E(Y|Z=0,A=1)}{P(c=co)/\{P(C=at)+P(C=co)\}}$ and $E(Y^0|C = co) = \frac{E(Y|Z=0,A=0) - [P(C=nt)/\{P(C=nt)+P(C=co)\}]E(Y|Z=1,A=0)}{P(c=co)/\{P(C=nt)+P(C=co)\}}$. The quantities $E(Y^1|C = nt)$ and $E(Y^0|C = at)$ are not identified. [2]

[24, 25, 26, 27] discuss using the framework described in this section to estimate the effect of a treatment from a randomized trial with non-compliance where random assignment is used as an IV. Several of these papers consider the additional complication of subsequent missing outcomes.

**Table 1.** The relation of observed groups and latent compliance classes under the monotonicity assumption

| $Z_i$ (observed) | $A_i$ (observed) | $C_i$ (latent) | | |
|---|---|---|---|---|
| 1 | 1 | $co$(Complier) | or | $at$ (Always-taker) |
| 1 | 0 | $nt$ (Never-taker) | | |
| 0 | 0 | $nt$ (Never-taker) | or | $co$ (Complier) |
| 0 | 1 | $at$ (Always-taker) | | |

## 3. Type of Unmeasured Confounding that Can Be Tested For Using an IV

In this section, we discuss what types of unmeasured confounding can be tested for using an IV. Because the concern about unmeasured confounding is that failure to adjust for unmeasured confounding might bias a treatment effect estimate, we define there to be unmeasured confounding for a treatment effect estimate if not adjusting for unmeasured variables would result in a biased estimate of the treatment effect[28]. There are different treatment effects that are of interest in causal inference, so it is possible for there to be unmeasured confounding for one type of treatment effect but not for another type of treatment effect.

We will consider testing for unmeasured confounding for three types of treatment effects that are often estimated in the causal inference literature. For ease of explanation, we will define the treatment effects conditional on the measured covariates $\mathbf{X}$; the treatment effect over the whole population of interest can be found by averaging the estimates of the treatment effect for different values of the measured covariates $\mathbf{X}$ over the covariate distribution of the population of interest[32]. The three treatment effects we will consider are the following: (i) the average causal effect over the whole population with covariates $\mathbf{X}$ (PACEX), $\tau_{pacex}(\mathbf{X}) = E(Y^1 - Y^0|\mathbf{X})$; (ii) the average causal effect for the treated subjects with covariates $\mathbf{X}$ (TACEX), $\tau_{tacex}(\mathbf{X}) = E(Y^1 - Y^0|A = 1, \mathbf{X})$; and (iii) the average causal effect for the compliers with covariates $\mathbf{X}$ (CACEX), $\tau_{cacex}(\mathbf{X}) = E(Y^1 - Y^0|C = co, \mathbf{X})$.

We say that there is no unmeasured confounding for a treatment effect $\tau(\mathbf{X})$ if the expectation of a straight comparison between the treated and control subjects' outcomes conditional on $\mathbf{X}$ is equal to $\tau(\mathbf{X})$, i.e.,

$$\text{No unmeasured confounding for } \tau(\mathbf{X}) : E[Y|A = 1, \mathbf{X}] - E[Y|A = 0, \mathbf{X}] = \tau(\mathbf{X}). \qquad (1)$$

When (1) holds, a consistent estimate of $\tau(\mathbf{X})$ under random sampling is the difference in sample means of the outcomes for treated ($A = 1$) and control ($A = 0$) subjects with covariates $\mathbf{X}$. When $\mathbf{X}$ is continuous or high dimensional, regression methods can be used to estimate $E[Y|A, \mathbf{X}]$, and $\tau(\mathbf{X})$ can be estimated by $\hat{E}[Y|A = 1, \mathbf{X}] - \hat{E}[Y|A = 0, \mathbf{X}]$ [32]. The bottom line is that when (1) holds, the treatment effect $\tau(\mathbf{X})$ can be estimated consistently from the observed data on treatments, outcomes and covariates; in contrast, when (1) does not hold, additional information like an IV is needed to deal with unmeasured confounding to obtain consistent estimates.

We now consider, under what conditions does no unmeasured confounding for the CACEX, $\tau_{cace}(\mathbf{X}) = E(Y^1 - Y^0|C = co)$, hold? No unmeasured confounding for the CACEX means

$$E[Y|A = 1, \mathbf{X}] - E[Y|A = 0, \mathbf{X}] = E[Y^1|C = co, \mathbf{X}] - E[Y^0|C = co, \mathbf{X}].$$

We can decompose the expected values on the left hand side of (1) into parts contributed by the compliance classes:

$$E(Y|A=1,\mathbf{X}) = E(Y^1|A=1,\mathbf{X}) = \frac{P(C=at|\mathbf{X})}{P(C=at|\mathbf{X})+P(Z=1|\mathbf{X})P(C=co|\mathbf{X})}E(Y^1|C=at,\mathbf{X})+$$
$$\frac{P(Z=1|\mathbf{X})P(C=co|\mathbf{X})}{P(C=at|\mathbf{X})+P(Z=1|\mathbf{X})P(C=co|\mathbf{X})}E(Y^1|C=co,\mathbf{X}),$$

$$E(Y|A=0,\mathbf{X}) = E(Y^0|A=0,\mathbf{X}) = \frac{P(C=nt|\mathbf{X})}{P(C=nt|\mathbf{X})+P(Z=0|\mathbf{X})P(C=co|\mathbf{X})}E(Y^0|C=nt,\mathbf{X})+$$
$$\frac{P(Z=0|\mathbf{X})P(C=co|\mathbf{X})}{P(C=nt|\mathbf{X})+P(Z=0|\mathbf{X})P(C=co|\mathbf{X})}E(Y^0|C=co,\mathbf{X}). \quad (2)$$

From (1) and (2), no unmeasured confounding for the CACEX holds if the following two conditions hold

$$E(Y^1|C=at,\mathbf{X}) = E(Y^1|C=co,\mathbf{X}) \quad (3)$$
$$E(Y^0|C=nt,\mathbf{X}) = E(Y^0|C=co,\mathbf{X}). \quad (4)$$

When (3) and (4) both hold, the CACEX can be estimated by comparing the treated and control groups conditional on $\mathbf{X}$ without using an IV. When (3) and/or (4) does not hold, then in most situations, estimating the CACEX by just comparing the treated and control groups conditional on $\mathbf{X}$ without using the IV will produce a biased estimate. Note that it is possible that such an estimate of the CACEX would be unbiased if, for example, the bias in $\hat{E}(Y|A=1,\mathbf{X})$ as an estimate of $E(Y^1|C=co,\mathbf{X})$ cancels the bias in $\hat{E}(Y|A=0,\mathbf{X})$ as an estimate of $E(Y^0|C=co,\mathbf{X})$, but such a situation is unlikely. Thus, for practical purposes, by testing (3) and (4), we are testing whether there is no unmeasured confounding for the CACEX. We will test (3) and (4) separately because the violation of one condition but not the other is informative about the nature of the unmeasured confounding; see the discussion of the NICU study in Section 7.

Because an IV only identifies the CACEX and does not identify the TACEX or the PACEX without further assumptions[2], we cannot test for no unmeasured confounding for the TACEX or PACEX without further assumptions. We now discuss further assumptions under which the test of (3)-(4) provides a test for no unmeasured confounding for the TACEX or PACEX. Consider first the TACEX. There is no unmeasured confounding for the TACEX if $E[Y|A=1,\mathbf{X}] - E[Y|A=0,\mathbf{X}] = E[Y^1 - Y^0|A=1,\mathbf{X}]$. If we have an assumption that guarantees

$$E[Y^1 - Y^0|A=1,\mathbf{X}] = E[Y^1 - Y^0|C=co,\mathbf{X}], \quad (5)$$

then no unmeasured confounding for the TACEX will be equivalent to no unmeasured confounding for the CACEX, and hence (3)-(4) will guarantee no unmeasured confounding for the TACEX. We will show that such an assumption that guarantees (5) is that the average treatment effect for the always takers and compliers is the same,

$$E(Y^1 - Y^0|C=at,\mathbf{X}) = E(Y^1 - Y^0|C=co,\mathbf{X}), \quad (6)$$

Under (6) and Assumptions 1-6 from Section 2.2, we have the following expression for $E(Y^1 - Y^0 | A = 1, \mathbf{X})$:

$$
\begin{aligned}
&E(Y^1 - Y^0 | A = 1, \mathbf{X}) \\
=\ &E(Y^1 - Y^0 | A = 1, C = co, \mathbf{X}) P(C = co | A = 1, \mathbf{X}) + E(Y^1 - Y^0 | C = at, \mathbf{X}) P(C = at | A = 1, \mathbf{X}) \\
=\ &E(Y^1 - Y^0 | Z = 1, C = co, \mathbf{X}) P(C = co | A = 1, \mathbf{X}) + E(Y^1 - Y^0 | C = at, \mathbf{X}) P(C = at | A = 1, \mathbf{X}) \\
=\ &E(Y^1 - Y^0 | C = co, \mathbf{X}) P(C = co | A = 1, \mathbf{X}) + E(Y^1 - Y^0 | C = at, \mathbf{X}) P(C = at | A = 1, \mathbf{X}) \\
=\ &E(Y^1 - Y^0 | C = co, \mathbf{X}),
\end{aligned}
\qquad (7)
$$

where the second to last equality follows from the assumption 2 that $Z$ is independent of potential outcomes and treatment received conditional on $\mathbf{X}$ and the last equality follows from (6) and the fact that

$$P(C = co | A = 1, \mathbf{X}) + P(C = at | A = 1, \mathbf{X}) = 1.$$

Thus, we conclude that under (6), the test of (3)-(4) tests whether there is no unmeasured confounding for the TACEX. Now consider the PACEX. If we have an assumption that guarantees

$$E[Y^1 - Y^0 | \mathbf{X}] = E[Y^1 - Y^0 | C = co, \mathbf{X}], \qquad (8)$$

then no unmeasured confounding for the PACEX will be equivalent to no unmeasured confounding for the CACEX. Using similar reasoning as in (7), such an assumption that guarantees (8) is that the average treatment effect for all three compliance classes – always takers, compliers and never takers – is the same,

$$E(Y^1 - Y^0 | C = at, \mathbf{X}) = E(Y^1 - Y^0 | C = co, \mathbf{X}) = E(Y^1 - Y^0 | C = nt, \mathbf{X}), \qquad (9)$$

Thus, under (9), the test of (3)-(4) tests whether there is unmeasured confounding for the PACEX. The assumption (9) means that the treatment effect is the same for the different compliance classes and thus that there is no effect modification by compliance class. No effect modification by compliance class will hold if either (a) treatment effects are homogeneous or (b) treatment effects are heterogeneous but the heterogeneity can be explained by observed covariates and we include those covariates in the covariate vector $\mathbf{X}$ that we condition on.

In summary, having an IV as defined in Section 2 enables us to test whether there is no unmeasured confounding for the average treatment effect for compliers (the CACE) by testing (3)-(4). An additional assumption of no effect modification by compliance class means that this test is also a test of whether there is no unmeasured confounding for the average treatment effect for the whole population.

In the NICU study (second to last paragraph of introduction and Section 7), there might be effect modification by compliance class since always takers tend to be sicker babies than compliers or never takers, and might benefit more from the high technology (e.g., sustained mechanical ventilation) in a high level NICU. Even if there is effect modification by compliance class, the test of (3)-(4) is still useful in two ways: (i) It tests for no unmeasured confounding for the CACE, which is useful to know about. The CACE is an important causal parameter because, by combining the CACE with information about how the treatment effect is expected to vary between compliers, always takers and never takers, we can predict the effects of increasing or decreasing access to the treatment[33, 34, 6], i.e., increasing or decreasing access to high level NICUs. If there is unmeasured confounding for the CACE, then the current study and future studies on the effect of high level NICUs on mortality and related outcomes should use IV methods to estimate the CACE. But if there is no unmeasured confounding, it is not necessary to use IV methods; (ii) If (3) or (4) is rejected, then that means there is an unmeasured variable, compliance class, which is associated with potential outcomes and treatment even after controlling for the observed covariates **X**. This suggests that some key covariate(s) relating to the treatment and outcome remains unmeasured and it would be worthwhile to try to collect more covariates relating to treatment and outcome if possible.

## 4. Compliance Class Model Test for Unmeasured Confounding Using an IV

Our approach to testing for no unmeasured confounding is to test (3)-(4) via the following approach:

1. Specify a model for the potential outcome and compliance class distributions.
2. Find the unconstrained maximum likelihood of the model using the EM algorithm.
3. Find the maximum likelihood of the model under the constraints (3) and (4) using the EM algorithm.
4. Test the validity of the constraints (3) and (4) using the likelihood ratio test.

We call our approach the *compliance class likelihood ratio test of no unmeasured confounding*. We will explain our approach in detail for binary outcomes and normally distributed outcomes.

*4.1. Binary Outcome Model*

We consider a logistic model for the outcome in each compliance class/treatment combination and a multinomial logistic model for the compliance classes. As a starting point, we consider a model in which the effect of the covariates **X** on the outcome is the same across the compliance classes.

*Model for the outcome*:

$$P(Y_i^a = 1 | C_i = t, \mathbf{X}_i = \mathbf{x}) = \frac{\exp(\kappa_{0,t} + \boldsymbol{\kappa}_1^T x + \gamma I(t = co)a)}{1 + \exp(\kappa_{0,t} + \boldsymbol{\kappa}_1^T x + \gamma I(t = co)a)}, \quad (10)$$

where $I(\cdot)$ is the indicator function. The $\kappa_{0,t}$'s measure the difference between the compliance classes when the IV $Z = 0$, $\boldsymbol{\kappa}_1$ is the effect of $x$, assumed to be the same across compliance classes and $\gamma$ is the log odds ratio for the effect of treatment for compliers.

*Model for compliance classes*: We consider a multinomial logit model. Let the compliers be the reference category.

$$
\begin{aligned}
P(C_i = co | \mathbf{X}_i = \mathbf{x}) &= \frac{1}{1 + \exp(\delta_{nt} + \boldsymbol{\tau}_{nt}^T \mathbf{x}) + \exp(\delta_{at} + \boldsymbol{\tau}_{at}^T \mathbf{x})} \\
P(C_i = nt | \mathbf{X}_i = \mathbf{x}) &= \frac{\exp(\delta_{nt} + \boldsymbol{\tau}_{nt}^T \mathbf{x})}{1 + \exp(\delta_{nt} + \boldsymbol{\tau}_{nt}^T \mathbf{x}) + \exp(\delta_{at} + \boldsymbol{\tau}_{at}^T \mathbf{x})} \\
P(C_i = at | \mathbf{X}_i = \mathbf{x}) &= \frac{\exp(\delta_{at} + \boldsymbol{\tau}_{at}^T \mathbf{x})}{1 + \exp(\delta_{nt} + \boldsymbol{\tau}_{nt}^T \mathbf{x}) + \exp(\delta_{at} + \boldsymbol{\tau}_{at}^T \mathbf{x})}.
\end{aligned}
\quad (11)
$$

We condition on the observed values of the covariates **X** and the IV $Z$, and seek to maximize the likelihood of $Y, A | Z, \mathbf{X}$ under the model (10)-(11). We can use the EM algorithm to maximize the likelihood of the model (10)-(11), where we think of the compliance classes as partially missing data as in ([7, 8, 35, 36]). If we know the compliance classes, the maximum likelihood estimate (MLE) is easy to compute. In practice, we are not able to observe the latent compliance classes, so we will use the EM algorithm to obtain the MLE. For the E step, conditional on observables and parameter estimates in the previous iteration, the expected value of the complete data log likelihood $l_C$ is

$$
\begin{aligned}
E(l_C|Y, Z, A, X) &= \sum_{i=1}^{n} I(Z_i = 1, A_i = 1) \log P(Z_i = 1|\mathbf{X}_i) \\
&+ I(Z_i = 1, A_i = 1)\omega_1[\log P(C_i = co|\mathbf{X}_i) + \log\{(P_i^{C1})^{y_i}\{1 - (P_i^{C1})\}^{1-y_i}\}] \\
&+ I(Z_i = 1, A_i = 1)(1 - \omega_1)[\log P(C_i = at|\mathbf{X}_i) + \log\{(P_i^{at})^{y_i}\{1 - (P_i^{at})\}^{1-y_i}\}] \\
&+ I(Z_i = 1, A_i = 0)[\log P(Z_i = 1|\mathbf{X}_i) + \log P(C_i = nt|\mathbf{X}_i) \\
&\quad + \log(P_i^{nt})^{y_i}\{1 - (P_i^{nt})\}^{1-y_i}] \\
&+ I(Z_i = 0, A_i = 0)\log\{1 - P(Z_i = 1|\mathbf{X}_i)\} \\
&+ I(Z_i = 0, A_i = 0)\omega_2[\log P(C_i = co|\mathbf{X}_i) + \log\{(P_i^{C0})^{y_i}\{1 - (P_i^{C0})\}^{1-y_i}\}] \\
&+ I(Z_i = 0, A_i = 0)(1 - \omega_2)[\log P(C_i = nt|\mathbf{X}_i)] + \log\{(P_i^{nt})^{y_i}\{1 - (P_i^{nt})\}^{1-y_i}\}] \\
&+ I(Z_i = 0, A_i = 1)[\log\{1 - P(Z_i = 1|\mathbf{X}_i)\} + \log P(C_i = at|\mathbf{X}_i) \\
&\quad + \log\{(P_i^{at})^{y_i}\{1 - (P_i^{at})\}^{1-y_i}\}]
\end{aligned}
$$

where

$$
\begin{aligned}
P_i^{C1} &= P(Y_i = 1|Z_i = 1, C_i = co, \mathbf{X}_i) = \frac{\exp(\kappa_{0,co} + \boldsymbol{\kappa}_1^T x + \gamma)}{1 + \exp(\kappa_{0,co} + \boldsymbol{\kappa}_1^T x + \gamma)} \\
P_i^{at} &= P(Y_i = 1|C_i = at, \mathbf{X}_i) = \frac{\exp(\kappa_{0,at} + \boldsymbol{\kappa}_1^T x)}{1 + \exp(\kappa_{0,at} + \boldsymbol{\kappa}_1^T x)} \\
P_i^{nt} &= P(Y_i = 1|C_i = nt, \mathbf{X}_i) = \frac{\exp(\kappa_{0,nt} + \boldsymbol{\kappa}_1^T x)}{1 + \exp(\kappa_{0,nt} + \boldsymbol{\kappa}_1^T x)} \\
P_i^{C0} &= P(Y_i = 1|Z_i = 0, C_i = co, \mathbf{X}_i) = \frac{\exp(\kappa_{0,co} + \boldsymbol{\kappa}_1^T x)}{1 + \exp(\kappa_{0,co} + \boldsymbol{\kappa}_1^T x)}
\end{aligned}
$$

$$
\begin{aligned}
\omega_1 &= P(C_i = co|Y_i = y, Z_i = A_i = 1, \mathbf{X}_i) \\
&= \frac{P(C_i = co|\mathbf{X}_i)f(Y_i|Z_i = 1, C_i = co, \mathbf{X}_i)}{P(C_i = co|\mathbf{X}_i)f(Y_i|Z_i = 1, C_i = co, \mathbf{X}_i) + P(C_i = at|\mathbf{X}_i)f(Y_i|Z_i = 1, C_i = at, \mathbf{X}_i)}
\end{aligned}
$$

$$
\begin{aligned}
\omega_2 &= P(C_i = co|Y_i = y, Z_i = A_i = 0, \mathbf{X}_i) \\
&= \frac{P(C_i = co|\mathbf{X}_i)f(Y_i|Z_i = 0, C_i = co, \mathbf{X}_i)}{P(C_i = co|\mathbf{X}_i)f(Y_i|Z_i = 0, C_i = co, \mathbf{X}_i) + P(C_i = nt|\mathbf{X}_i)f(Y_i|Z_i = 0, C_i = nt, \mathbf{X}_i)}
\end{aligned}
$$

We seek to test the null hypothesis that (3) and (4) hold. Under the model (10)-(11), these two equations can be simplified as

$$\frac{\exp(\kappa_{0,at} + \boldsymbol{\kappa}_1^T x)}{1 + \exp(\kappa_{0,at} + \boldsymbol{\kappa}_1^T x)} = \frac{\exp(\kappa_{0,co} + \boldsymbol{\kappa}_1^T x + \gamma)}{1 + \exp(\kappa_{0,co} + \boldsymbol{\kappa}_1^T x + \gamma)};$$

$$\frac{\exp(\kappa_{0,co} + \boldsymbol{\kappa}_1^T x)}{1 + \exp(\kappa_{0,co} + \boldsymbol{\kappa}_1^T x)} = \frac{\exp(\kappa_{0,nt} + \boldsymbol{\kappa}_1^T x)}{1 + \exp(\kappa_{0,nt} + \boldsymbol{\kappa}_1^T x)}.$$

By simple calculation, the null hypothesis that (3) and (4) hold is equivalent to the constraints

$$
\begin{aligned}
\kappa_{0,at} &= \kappa_{0,co} + \gamma, \\
\kappa_{0,co} &= \kappa_{0,nt}.
\end{aligned}
\tag{12}
$$

Under the null hypothesis, we can maximize the likelihood using the EM algorithm, where we impose the constraints (12). The likelihood ratio test of (3) and (4) looks at the difference between the maximized log likelihood of the unconstrained model (10)-(11) and the maximized log likelihood of the model that constrains (3) and (4) to hold, i.e., (12) to hold; under the null hypothesis, 2 times this difference follows asymptotically a $\chi^2$ distribution with degrees of freedom equal to the number of extra free parameters in the unconstrained model compared to the constrained model [37]. For the model (10)-(11), the constraints (12) reduce the number of free parameters by 2 (since $\kappa_{0,at}$ and $\kappa_{0,nt}$ are no longer free parameters given $\kappa_{0,co}$ and $\gamma$). Note that for the likelihood ratio test, we are considering the maximized observed data log likelihood rather than the complete data log likelihood. We can test (3) and (4) separately by carrying out likelihood ratio tests of the constraints $\kappa_{0,at} = \kappa_{0,co} + \gamma$ and $\kappa_{0,co} = \kappa_{0,nt}$; for each of these tests, the null distribution of 2 times the log likelihood ratio is $\chi^2$ with 1 degree of freedom.

We now consider a binary outcome model where the treatment effect is heterogeneous, that is, the treatment effect depends on covariates $\mathbf{X}$.

*Model for the outcome that allows for heterogeneous treatment effects*: With covariates $\mathbf{X}$, we consider the following model. Let $\lambda_{t,\mathbf{x}}$ denote the log odds that $y = 1$ when $z = 0$ for compliance class $t$ and covariate vector $\mathbf{x}$. Let $\gamma_{\mathbf{x}}$ be the log odds ratio that $y = 1$ for $A = 1$ vs. $A = 0$ for compliers with covariates $\mathbf{X}$. The model we consider is

$$\begin{aligned}
P(Y_i^a = 1 | C_i = t, \mathbf{X}_i = \mathbf{x}) &= \frac{\exp(\lambda_{t,\mathbf{x}} + \gamma_{\mathbf{x}} I(t = co)a)}{1 + \exp(\lambda_{t,\mathbf{x}} + \gamma_{\mathbf{x}} I(t = co)a)} \\
\gamma_{\mathbf{x}} &= \alpha_0 + \boldsymbol{\alpha}_1^T \mathbf{x} \\
\lambda_{t,\mathbf{x}} &= \kappa_{0,t} + \boldsymbol{\kappa}_{1,t}^T \mathbf{x}.
\end{aligned} \quad (13)$$

The model allows for the treatment effect for compliers to depend on $\mathbf{x}$ through $\boldsymbol{\alpha}_1$ and for the difference between the compliance classes when $Z = 0$ to depend on $\mathbf{x}$ through the $\boldsymbol{\kappa}_{1,t}$'s.

The expected value of the complete data log likelihood $l_C$ and the observed data log likelihood for the binary outcome model with heterogeneous treatment effect (13) is of the same form as those with homogeneous treatment effect with different expressions for $P_i^{C1}, P_i^{C0}, P_i^{at}$ and $P_i^{nt}$:

$$\begin{aligned}
P_i^{C1} &= P(Y_i = 1 | Z_i = 1, C_i = co, \mathbf{X}_i) = \frac{\exp(\lambda_{co,\mathbf{x}} + \gamma_{\mathbf{x}})}{1 + \exp(\lambda_{co,\mathbf{x}} + \gamma_{\mathbf{x}})} \\
P_i^{at} &= P(Y_i = 1 | C_i = at, \mathbf{X}_i) = \frac{\exp(\lambda_{at,\mathbf{x}})}{1 + \exp(\lambda_{at,\mathbf{x}})} \\
P_i^{nt} &= P(Y_i = 1 | C_i = nt, \mathbf{X}_i) = \frac{\exp(\lambda_{nt,\mathbf{x}})}{1 + \exp(\lambda_{nt,\mathbf{x}})} \\
P_i^{C0} &= P(Y_i = 1 | Z_i = 0, C_i = co, \mathbf{X}_i) = \frac{\exp(\lambda_{co,\mathbf{x}})}{1 + \exp(\lambda_{co,\mathbf{x}})}
\end{aligned}$$

By simple calculation, the null hypothesis of no unmeasured confounding (3)-(4) under model (13) is

$$H_0: \quad \boldsymbol{\kappa}_{nt} = \boldsymbol{\kappa}_{co},$$
$$\boldsymbol{\kappa}_{at} = \boldsymbol{\kappa}_{co} + \boldsymbol{\alpha}, \quad (14)$$

where $\boldsymbol{\kappa}_t = (\kappa_{0,t}, \boldsymbol{\kappa}_{1,t}^T)$ and $\boldsymbol{\alpha} = (\alpha_0, \boldsymbol{\alpha}_1)^T$.

*4.2. Normal Outcome Model*

A normal model for the outcome that allows for a heterogeneous treatment effect that is analogous to (13) is

$$f(Y_i^a|C_i = t, \mathbf{X}_i = \mathbf{x}) = N(\kappa_{0,t} + \boldsymbol{\kappa}_{1,t}^T\mathbf{x} + (\alpha_0 + \boldsymbol{\alpha}_1^T\mathbf{x})I(t = co)a, \sigma^2) \quad (15)$$

The null hypothesis of no unmeasured confounding (3)-(4) under the model (15) is (14).

*4.3. Computation of MLE– EM and BFGS Optimization*

The EM algorithm can sometimes be slow to converge to the MLE near the maximizer of the likelihood[39, 40]. To speed up the convergence to the MLE as in [39, 40, 41], we first run the EM algorithm until it comes close to convergence and then use the EM estimates as the starting value and maximize the likelihood by a quasi-Newton method, the Broyden-Fletcher-Goldfarb-Shanno algorithm (BFGS) as implemented in the optim function in R[42]. R functions for computing the EM estimates and BFGS method and implementing our compliance class likelihood ratio test of no unmeasured confounding are provided in the supplementary materials. Instructions for using the functions and an example data set are also provided.

## 5. DWH Test

In this section, we will consider the DWH test statistic for testing for no unmeasured confounding using an IV and its properties. The conventional DWH test is formulated for a model with a continuous outcome[10, 11, 12]. The DWH test statistic $T_{DWH}$ is the following. Let $\hat{\beta}_{OLS}$ denote the ordinary least squares (OLS) estimate of the effect of $A$ on $Y$ controlling for $\mathbf{X}$. Let $\hat{\beta}_{2SLS}$ denote the two stage least squares estimate of the effect of $A$ on $Y$ controlling for $\mathbf{X}$ using $Z$ as an IV; $\hat{\beta}_{2SLS}$ is computed by by first regressing $A$ on $Z, \mathbf{X}$ by least squares and finding the predicted value $\hat{A}$ and then regressing $Y$ on $\hat{A}, \mathbf{X}$ by least squares. The DWH test statistic is an assessment of the difference between the OLS and 2SLS estimates of the causal effect of $A$ on $Y$,

$$T_{DWH} = \frac{(\hat{\beta}_{OLS} - \hat{\beta}_{2SLS})^2}{\hat{Var}(\hat{\beta}_{2SLS}) - \hat{Var}(\hat{\beta}_{OLS})}. \quad (16)$$

[43]. The variances in (16) are the variances that come from the normal linear regression model and the normal simultaneous equations model that make the homoskedasticity assumption that $Var(Y^0|\mathbf{X})$ is equal to $Var(Y^1|\mathbf{X})$ and the same for all $\mathbf{X}$[43]. Note that there are several asymptotically equivalent forms of the DWH test which differ in the way the denominator in (16) is computed; see [43], pp. 50-52. The null hypothesis of the DWH test can be expressed as the following: there is no unmeasured confounding, i.e., (3) and (4) hold, and the following three assumptions hold: (i) the average treatment effect for compliers is homogeneous in $\mathbf{X}$, i.e., $E(Y^1 - Y^0|C = co, \mathbf{X})$ is the same for all $\mathbf{X}$; (ii) $E(Y^0|\mathbf{X}))$ is linear in $\mathbf{X}$; and (iii) a homoskedasticity assumption that $Var(Y^0|\mathbf{X})$ is equal to $Var(Y^1|\mathbf{X})$ and the same for all $\mathbf{X}$. The asymptotic null distribution of $T_{DWH}$ is chi-squared with 1 degree of freedom [10, 12, 11, 43]. Under the null hypothesis for the DHW test, the denominator of the DHW test statistic (16) times the sample size $N$ converges to the variance of $\sqrt{N}$ times $(\hat{\beta}_{OLS} - \hat{\beta}_{2SLS})$ [12].

We now consider the properties of the DWH test when average treatment effects for compliers are heterogeneous in $\mathbf{X}$. When average treatment effects for compliers are heterogeneous in $\mathbf{X}$, the DWH test may reject with probability converging to 1 even when there is no unmeasured confounding. To show this, we will show that $\hat{\beta}_{OLS}$ and $\hat{\beta}_{TSLS}$ can converge to different weighted averages of treatment effects when average treatment effects for compliers are heterogeneous in $\mathbf{X}$. Combining this fact with the fact that, under regularity conditions described in [44], the denominator of the DWH test statistic (16) will converge to $0$, shows that $T_{DWH}$ converges in probability to $\infty$ and rejects with probability $1$ even when there is no unmeasured confounding.

We now consider the properties of $\hat{\beta}_{OLS}$ when (3)-(4) hold and treatment effects are heterogeneous in $\mathbf{X}$. Let $\beta_{\mathbf{X}} = E(Y^1 - Y^0 | C = co, \mathbf{X})$. Suppose $E(Y_i | \mathbf{X}_i, A_i = 0)$ is linear in $\mathbf{X}_i$. Then $E(Y|A=1, \mathbf{X}) - E(Y|A=0, \mathbf{X}) = \beta_{\mathbf{X}}$. Then, under the assumption that (3)-(4) hold, we have the following expression for the probability limit of the OLS estimator where $E^*(A|\mathbf{B})$ is the linear projection of $A$ onto $\mathbf{B}$ (i.e., $E^*(A|\mathbf{B}) = \boldsymbol{\alpha}^T \mathbf{B}$, $\boldsymbol{\alpha} = \arg\min_{\boldsymbol{\alpha}^*} E(A - (\boldsymbol{\alpha}^*)^T \mathbf{B}))$,

$$plim\hat{\beta}_{OLS} = \frac{E[(A_i - E^*(A_i|\mathbf{X}_i))(Y_i - E^*(Y_i|\mathbf{X}_i))]}{E[(A_i - E^*(A_i|\mathbf{X}_i))^2]} \quad (17)$$

$$= \frac{E[(A_i - E^*(A_i|\mathbf{X}_i))Y_i]}{E[(A_i - E^*(A_i|\mathbf{X}_i))^2]} \quad (18)$$

$$= \frac{E[(A_i - E^*(A_i|\mathbf{X}_i))E(Y_i|\mathbf{X}_i, A_i)]}{E[(A_i - E^*(A_i|\mathbf{X}_i))^2]} \quad (19)$$

$$= \frac{E[(A_i - E^*(A_i|\mathbf{X}_i))(E(Y_i|\mathbf{X}_i, A_i = 0) + \beta_{\mathbf{X}} A_i)]}{E[(A_i - E^*(A_i|\mathbf{X}_i))^2]} \quad (20)$$

$$= \frac{E[(A_i - E^*(A_i|\mathbf{X}_i))^2 \beta_{\mathbf{X}}]}{E[(A_i - E^*(A_i|\mathbf{X}_i))^2]}, \quad (21)$$

where we used the fact that $E((A_i - E^*(A_i|\mathbf{X}_i))X_i) = 0$ to derive (18) and (21) and we iterated expectations over $\mathbf{X}_i$ and $A_i$ to derive (19). (21) shows that the OLS estimator converges to a weighted average of treatment effects at different values of $\mathbf{X}$, where the values of $\mathbf{X}$ that get the most weight are those where $E[(A_i - E^*(A_i|\mathbf{X}_i))^2]$ is largest. [38] derive similar expressions assuming $E(A_i|\mathbf{X}_i)$ is linear in $\mathbf{X}_i$. If $E(A_i|\mathbf{X}_i)$ is linear in $\mathbf{X}$, then $E[(A_i - E^*(A_i|\mathbf{X}_i))^2]$ is the conditional variance of $A$ given $\mathbf{X}$. If $E(Y_i|\mathbf{X}_i, A_i = 0)$ is not linear in $\mathbf{X}_i$, then $plim\hat{\beta}_{OLS}$ equals (21) plus $\frac{E[(A_i - E^*(A_i|\mathbf{X}_i))E(Y^0|\mathbf{X})]}{E[(A_i - E^*(A_i|\mathbf{X}_i))^2]}$.

We now consider the properties of $\hat{\beta}_{TSLS}$ when (3)-(4) hold but treatment effects are heterogeneous in $\mathbf{X}$. We assume $E(A|\mathbf{X}, Z)$ is linear in $\mathbf{X}, Z$. Then the $plim$ of $\hat{\beta}_{TSLS}$ is the $plim$ of the coefficient on $E(A|\mathbf{X}, Z)$ in the regression of $Y$ on $E(A|\mathbf{X}, Z)$ and $\mathbf{X}$. By the same reasoning as in (17)-(21), this $plim$ is the weighted average of $\beta_{\mathbf{X}}$ over the distribution of $\mathbf{X}$, weighted by $E[\{E(A|\mathbf{X}, Z) - E^*(E(A|\mathbf{X}, Z)|\mathbf{X})\}^2]$. Under the assumption that $E(A|\mathbf{X}, Z)$ is linear in $\mathbf{X}, Z$, these weights equal the conditional variance of $E(A|Z, \mathbf{X})$ given $\mathbf{X}$, which equals $P(C = co|\mathbf{X})^2 P(Z = 1|\mathbf{X})(1 - P(Z = 1|\mathbf{X}))$. Thus,

$$plim\hat{\beta}_{TSLS} = E[P(C = co|\mathbf{X}_i)^2 P(Z_i = 1|\mathbf{X}_i)(1 - P(Z_i = 1|\mathbf{X}_i))\beta_{\mathbf{X}}]/$$
$$E[P(C = co|\mathbf{X}_i)^2 P(Z_i = 1|\mathbf{X}_i)(1 - P(Z_i = 1|\mathbf{X}_i))]. \quad (22)$$

Thus, the TSLS estimator converges to a weighted average of treatment effects at different values of $\mathbf{X}$, where the values of $\mathbf{X}$ that tend to get the most weight are those for which the proportion of compliers is highest. A value of $\mathbf{X}$ at which there are no compliers gets zero weight.

From (21) and (22), the numerator of the DWH test statistic (16) converges in probability to

$$\frac{E[P(C = co|\mathbf{X})^2 P(Z = 1|\mathbf{X})(1 - P(Z = 1|\mathbf{X}))\beta_{\mathbf{X}}]}{E[P(C = co|\mathbf{X})^2 P(Z = 1|\mathbf{X})(1 - P(Z = 1|\mathbf{X}))]} - \frac{E[(A - E^*(A|\mathbf{X}))^2 \beta_{\mathbf{X}}]}{E[(A - E^*(A|\mathbf{X}))^2]}. \quad (23)$$

and the denominator of (16) converges in probability to $0$ under regularity conditions. Thus, under the regularity conditions, when (23) is not equal to zero, $T_{DWH}$ converges in probability to $\infty$. In summary, even when there is no unmeasured confounding, we have shown that $\hat{\beta}_{OLS}$ and $\hat{\beta}_{TSLS}$ can converge to different weighted averages of treatment effects when average treatment effects for compliers are heterogeneous in $\mathbf{X}$, and consequently the DWH test statistic (16) can converge in probability to $\infty$.

## 6. Simulation Study

*6.1. Normal Outcomes*

We will compare the compliance class likelihood ratio test of no unmeasured confounding developed in Section 4 to the DWH test in a simulation study under the normal outcome model (15). We will consider one binary covariate $X$. We consider three scenarios:

I The null hypothesis of no unmeasured confounding for the CACE, i.e., (3)-(4), holds, which is equivalent for the normal outcome model to (14) holding. Additionally, the complier treatment effect is homogeneous in $X$, i.e., $\alpha_1 = 0$. Here we expect that both the DWH test and our test will have a $0.05$ Type I error rate.

II The null hypothesis of no unmeasured confounding for the CACE holds but the treatment effect is heterogeneous in $X$, i.e., $\alpha_1 \neq 0$. Here we expect our test will have a $0.05$ Type I error rate but the DWH test will have a greater than $0.05$ Type I error rate.

III The null hypothesis of no unmeasured confounding for the CACE does not hold.

The parameters for (15) for each scenario are shown in Table 2. The sample size for each simulation scenario is 1000 and 1000 simulations were carried out for each scenario. For all the scenarios, the IV $Z$ was generated as a Bernoulli random variable with

$$P(Z_i = 1 | X_i = x) = \frac{\exp(-1 + 2x)}{1 + \exp(-1 + 2x)}.$$

The parameters $-1$ and $2$ were chosen so that $P(Z|\mathbf{X})$ and the marginal probability of $Z = 1$ is about $\frac{1}{2}$. Also, for all the scenarios, the model for the compliance class is as follows:

$$P(C_i = at | X_i = x) = P(C_i = nt | X_i = x) = \frac{\exp(-2.5 + 3.5x)}{1 + 2\exp(-2.5 + 3.5x)}$$
$$P(C_i = co | X_i = x) = \frac{1}{1 + 2\exp(-2.5 + 3.5x)} \quad (24)$$

From section 5, we know that the OLS estimator is a weighted average of the treatment effects with conditional variance $Var(A_i|X_i)$ as weights whereas the 2SLS estimator is a weighted average of the treatment effects where the weights are related with $P(C_i = co|X_i)$. The compliance class model (24) was chosen so that these two sets of weights differ. In scenario II, we have $P(C_i = co|X_i = 1) \approx 0.16$, $P(C_i = co|X_i = 0) \approx 0.85$ whereas $Var(A_i|X_i = 1) \approx 0.25$, $Var(A_i|X_i = 0) \approx 0.22$.

Table 3 shows the bias of the MLE estimates over the whole parameter space for the three simulation scenarios as well as the bias of the restricted MLE estimates (RMLE) under the constraint (14) of no unmeasured confounding for the CACE. The MLE estimates are approximately unbiased for all three scenarios. The RMLE estimates are approximately unbiased for the first two scenarios where no unmeasured confounding for the CACE holds, but are substantially biased in Scenario 3, where no unmeasured confounding for the CACE does not hold.

Table 4 shows the rejection rates of the compliance class model likelihood ratio test and the DWH test for the three scenarios. For scenario I, in which the null hypothesis of no unmeasured confounding for the CACE holds and also the treatment effect for compliers is homogeneous in $X$, both the compliance class model likelihood ratio test and the DWH test have empirical rejection rates close to their nominal Type I error level. For scenario II, in which the null hypothesis of no unmeasured confounding for the CACE holds but the treatment effect for compliers is heterogeneous in $X$, the compliance class model likelihood ratio test has an empirical rejection rate close to its nominal Type I error rate but the DWH test rejects far too often, e.g., it rejects 0.853 of the time when the nominal type I error rate is 0.05. For scenario III, in which the null hypothesis of no unmeasured confounding for the CACE does not hold, both tests have high power with the compliance class model likelihood ratio test having slightly higher power. In summary, the simulation study results show that the compliance class model likelihood ratio test has advantages over the DWH test: the compliance class model likelihood ratio test has comparable power to the DWH test, but keeps close to the correct Type I error rate when treatment effects are heterogeneous in $X$, unlike the DWH test.

|  | Scenario I | Scenario II | Scenario III |
|---|---|---|---|
| $\kappa_{0,at}$ | 0.8 | 0.8 | 1.5 |
| $\kappa_{1,at}$ | 1 | 0 | 1 |
| $\kappa_{0,co}$ | 0.3 | 0.3 | 0.3 |
| $\kappa_{1,co}$ | 1 | 1 | 1 |
| $\kappa_{0,nt}$ | 0.3 | 0.3 | -1 |
| $\kappa_{1,nt}$ | 1 | 1 | 2 |
| $\alpha_0$ | 0.5 | 0.5 | 0.5 |
| $\alpha_1$ | 0 | -1 | -1 |

**Table 2.** Parameters of Normal Outcome Model

|  | Scenario I | | Scenario II | | Scenario III | |
|---|---|---|---|---|---|---|
|  | MLE | RMLE | MLE | RMLE | MLE | RMLE |
| $\delta_{nt}$ | -0.0477146 | -0.0544274 | -0.0476901 | -0.0543439 | 0.0156288 | -0.0543725 |
| $\tau_{nt}$ | 0.1013498 | 0.0999249 | 0.1007317 | 0.0997708 | 0.0128341 | 0.0997589 |
| $\delta_{at}$ | -0.0095424 | -0.0104721 | -0.0095449 | -0.0104576 | 0.0038682 | -0.0104442 |
| $\tau_{at}$ | 0.0519741 | 0.0442860 | 0.0513301 | 0.0441942 | 0.0009426 | 0.0441336 |
| $\kappa_{0,at}$ | 0.0058613 | -0.0005772 | 0.0058594 | -0.0005727 | 0.0037276 | -0.5367873 |
| $\kappa_{1,at}$ | -0.0062825 | 0.0006663 | -0.0063618 | 0.0006561 | 0.0014474 | 0.1747882 |
| $\kappa_{0,co}$ | -0.0018516 | -0.0020254 | -0.0018565 | -0.0020229 | 0.0006294 | -0.1324547 |
| $\kappa_{1,co}$ | -0.0166094 | 0.0037380 | -0.0142488 | 0.0037335 | 0.0596318 | -0.1388161 |
| $\kappa_{0,nt}$ | -0.0031835 | -0.0021490 | -0.0031585 | -0.0021505 | 0.0111069 | 1.1674984 |
| $\kappa_{1,nt}$ | 0.0066654 | 0.0039132 | 0.0064480 | 0.0039194 | 0.0093282 | -1.1386382 |
| $\alpha_0$ | -0.0010242 | 0.0016277 | -0.0010176 | 0.0016200 | 0.0069120 | 0.2959712 |
| $\alpha_1$ | 0.0326921 | -0.0031748 | 0.0307796 | -0.0031632 | 0.0566909 | 1.3134260 |
| $\sigma^2$ | -0.0135517 | -0.0025765 | -0.0135338 | -0.0025765 | 0.0079252 | 0.0920937 |

**Table 3.** Bias of Sample mean of $\delta_t$, $\tau_t$, $\lambda_t$, $\gamma_t$ and $\sigma^2$ with 1000 simulations of sample size 1000 for the normal model

| Scenario | I | | II | | III | |
|---|---|---|---|---|---|---|
|  | $\widehat{rr}_{\alpha=0.01}$ | $\widehat{rr}_{\alpha=0.05}$ | $\widehat{rr}_{\alpha=0.01}$ | $\widehat{rr}_{\alpha=0.05}$ | $\widehat{rr}_{\alpha=0.01}$ | $\widehat{rr}_{\alpha=0.05}$ |
| DWH test | 0.006 | 0.042 | 0.662 | 0.853 | 0.98 | 0.997 |
| LR test | 0.014 | 0.053 | 0.014 | 0.052 | 1 | 0.999 |

**Table 4.** Empirical rejection rates $\widehat{rr}_\alpha$ with type I error $\alpha$ for the normal outcome model with 1000 simulations.

The compliance class model likelihood ratio test (the LR test) is more computationally intensive than the DWH test. For the first scenario of the simulation study, on an Optiplex 780 computer with Intel Core Duo CPU E8400 @ 3.00 GHZ and 4 GB Ram, the DWH test took an average of 0.013 CPU seconds while the LR test took an average of 53.7 CPU seconds. The DWH test is faster than the compliance class likelihood ratio test, but the LR test is not prohibitively slow, taking less than a minute. Although the DWH test is faster, the DWH test does not work for models with heterogeneous treatment effects. When dealing with real data, we do not know whether there are heterogeneous treatment effects. When computational efficiency is not the main concern, we suggest to test for no unmeasured confounding using the LR test.

*6.2. Binary Outcomes*

In this section, we study the performance of the compliance class likelihood ratio for no unmeasured confounding for binary outcomes developed in Section 4.1 We do not consider the DWH since the DWH test assumes normal outcomes rather than binary outcomes. We simulate the binary outcome model as in (10) and estimate the parameters by the EM algorithm for the binary outcome. As for the simulation for normal outcomes, we consider three scenarios: (I) the null hypothesis of no unmeasured confounding for the CACE holds and the complier treatment effect is homogeneous in $X$; (II) the null hypothesis of no unmeasured confounding for the CACE holds and the complier treatment effect is heterogeneous in $X$; and (III) the null hypothesis of no unmeasured confounding for the CACE does not hold. The parameters for (10) for each scenario are shown in Table 5. The sample size for each simulation scenario is 1000 and 1000 simulations were carried out for each scenario. For all the scenarios, the IV $Z$ was generated as a Bernoulli random variable with

$$P(Z_i = 1 | X_i = x) = \frac{\exp(-2+x)}{1+\exp(-2+x)}.$$

The parameters $-2$ and $1$ were chosen so that $P(Z|\mathbf{X})$ and the marginal probability of $Z=1$ is about $\frac{1}{2}$. Also, for all the scenarios, the model for the compliance class is as follows:

$$\begin{aligned}
P(C_i = at | X_i = x) &= \frac{\exp(-1.5+0.1x)}{1+\exp(-1.5+0.1x)+\exp(-1+0.05x)} \\
P(C_i = nt | X_i = x) &= \frac{\exp(-1+0.05x)}{1+\exp(-1.5+0.1x)+\exp(-1+0.05x)} \\
P(C_i = co | X_i = x) &= \frac{1}{1+\exp(-1.5+0.1x)+\exp(-1+0.05x)}
\end{aligned} \qquad (25)$$

Table 6 shows the bias of the MLE estimates over the whole parameter space for the three simulation scenarios as well as the bias of the restricted MLE estimates (RMLE) under the constraint (14) of no unmeasured confounding for the CACE. The MLE estimates are approximately unbiased for all three scenarios. The RMLE estimates are approximately unbiased for the first two scenarios where no unmeasured confounding for the CACE holds, but are substantially biased in Scenario 3, where no unmeasured confounding for the CACE does not hold.

Table 7 shows the rejection rates of the compliance class model likelihood ratio test for the three scenarios. For scenario I and scenario II, in which the null hypothesis of no unmeasured confounding for the CACE holds, the compliance class model likelihood ratio test has an empirical rejection rate close to its nominal Type I error rate. For scenario III, in which the null hypothesis of no unmeasured confounding for the CACE does not hold, the compliance class model likelihood ratio test has high power.

|  | Scenario I | Scenario II | Scenario III |
|---|---|---|---|
| $\kappa_{0,at}$ | 0 | 0 | -2 |
| $\kappa_{1,at}$ | 1 | 0 | 1 |
| $\kappa_{0,co}$ | -0.5 | -0.5 | -0.5 |
| $\kappa_{1,co}$ | 1 | 1 | 1 |
| $\kappa_{0,nt}$ | -0.5 | -0.5 | -1 |
| $\kappa_{1,nt}$ | 1 | 1 | 0.5 |
| $\alpha_0$ | 0.5 | 0.5 | 0.5 |
| $\alpha_1$ | 0 | -1 | -1 |

**Table 5.** Parameters of Binary Outcome Model

|  | Scenario I | | Scenario II | | Scenario III | |
|---|---|---|---|---|---|---|
|  | MLE | RMLE | MLE | RMLE | MLE | RMLE |
| $\delta_{nt}$ | -0.009862 | -0.010882 | -0.005490 | -0.006288 | 0.000971 | -0.004599 |
| $\tau_{nt}$ | 0.003475 | 0.003582 | 0.003023 | 0.002972 | -0.000865 | 0.001501 |
| $\delta_{at}$ | -0.009214 | -0.009853 | -0.002856 | -0.003171 | -0.001430 | -0.006231 |
| $\tau_{at}$ | -0.001601 | -0.001790 | 0.001344 | 0.000897 | -0.004487 | -0.001151 |
| $\kappa_{0,at}$ | -0.025060 | -0.015066 | -0.005103 | -0.005550 | -0.125930 | 1.268413 |
| $\kappa_{1,at}$ | 0.074988 | 0.013357 | 0.008522 | 0.001488 | 0.076193 | -0.718085 |
| $\kappa_{0,co}$ | -0.032044 | -0.007961 | -0.043709 | -0.011160 | -0.028737 | -0.116232 |
| $\kappa_{1,co}$ | 0.046542 | 0.008069 | 0.052011 | 0.009544 | 0.046948 | -0.419943 |
| $\kappa_{0,nt}$ | -0.088219 | -0.007961 | -0.086033 | -0.011160 | -0.033784 | 0.383768 |
| $\kappa_{1,nt}$ | 0.066193 | 0.008069 | 0.067434 | 0.009544 | 0.017085 | 0.080057 |
| $\alpha_0$ | 0.032538 | -0.007106 | 0.035052 | 0.005610 | 0.018919 | -0.615355 |
| $\alpha_1$ | 0.098868 | 0.005288 | -0.052364 | -0.008057 | -0.047160 | 0.701858 |

**Table 6.** Bias of Sample mean of $\delta_t$, $\tau_t$, $\lambda_t$, $\gamma_t$ and $\sigma^2$ with 1000 simulations of sample size 1000 for the binary outcome model

| Scenario | I | | II | | III | |
|---|---|---|---|---|---|---|
|  | $\widehat{rr}_{\alpha=0.01}$ | $\widehat{rr}_{\alpha=0.05}$ | $\widehat{rr}_{\alpha=0.01}$ | $\widehat{rr}_{\alpha=0.05}$ | $\widehat{rr}_{\alpha=0.01}$ | $\widehat{rr}_{\alpha=0.05}$ |
| LR test | 0.050 | 0.014 | 0.013 | 0.055 | 0.994 | 0.997 |

**Table 7.** Empirical rejection rates $\widehat{rr}_\alpha$ with type I error $\alpha$ for the exponential(1) distribution with 1000 simulations.

## 7. Application to Study of High-Level NICUs vs. Lower-Level NICUs

We obtained birth certificates from all deliveries occurring in Pennsylvania between 1995-2005. The Pennsylvania Department of Health linked these birth certificates to death certificates using name and date of birth, and then de-identified the records. We then matched over 98% of birth certificates to maternal and newborn hospital records using methods described in [21]. Over 80% of the unmatched birth certificate records were missing hospital, suggesting a birth at home or a birthing center. The unmatched records had similar gestational age and racial/ethnic distributions to the matched records. The Institutional Review Boards of The Children's Hospital of Philadelphia and the Pennsylvania Department of Health approved this study.

Infants included in this study had a gestational age between 23 and 37 weeks, and a birth weight between 400 to 8000 grams. Birth records were excluded if the birth weight was more than 5 standard deviations from the mean birth weight for the recorded gestational age in the cohort. There are 192,078 infants in the final cohort. The primary outcome for this study is neonatal death, defined as any death during the initial birth hospitalization. The IV we consider is $Z = 1$ if a mother's excess travel time to the nearest high level NICU compared to the nearest hospital is 10 minutes or less, $Z = 0$ if her excess travel time is more than 10 minutes. The measured confounders $\mathbf{X}$ are birth weight, an indicator for whether birth weight is missing, the month of pregnancy that prenatal care started, an indicator for whether this month is missing, mother's education, an indicator for whether mother's education is missing and an indicator for whether the mother went into labor prematurely (as compared to having a planned premature birth by induced labor or C-section).

Mother's excess travel time $Z$ plausibly satisfies the IV assumptions 1-6 in Section 2.2 for the following reasons. First, for the stable unit treatment value assumption (SUTVA), whether one mother lives near a high level NICU or delivers at a high level NICU is unlikely to affect another mother and her infant, so SUTVA is plausible. Second, for the IV being independent of unmeasured confounding, women do not expect to have a premature delivery, and thus conditional on measured socioeconomic variables such as mother's education, women do not choose where to live based on distance to a high level NICU, making independence from unmeasured confounding plausible[21]. Third, the exclusion restriction (no direct effect of excess travel time) is plausible because most mothers have time to deliver at either the nearest high level or low level NICU so that the marginal travel time should not directly affect outcomes[21]. Fourth, for the IV affecting the treatment, excess travel time is correlated with whether a mother delivers at a high level NICU because a mother typically obtains prenatal care from and would prefer to deliver at a close by facility[23]. Fifth, for the monotonicity assumption, if a mother would travel to go to a high level NICU when living more than ten minutes further from the high level NICU than the nearest low level NICU, she would presumably also travel to the high level NICU if it were less than ten minutes further than the nearest low level NICU; thus, monotonicity is plausible. Sixth, for the consistency assumption, although it is unlikely to hold exactly since different high level NICUs may differ in their level of care and different low level NICUs may differ in their level of care, it plausibly holds approximately; see [45] for discussion about interpreting causal estimates when consistency does not hold exactly. In summary, mother's excess travel time is a plausible IV.

[21] and [9] have used mother's excess travel time as an IV to estimate the effect of a premature infant being delivered in a high level NICU vs. a low level NICU for compliers, i.e., $E(Y^1 - Y^0 | C = co)$. Here our focus is on using mother's excess travel time to test whether there is unmeasured confounding. Because the outcome, neonatal death, is binary, we consider the binary outcome models of Section 4.1. First, we test the null hypothesis that there are homogeneous treatment effects in terms of the measured covariates $\mathbf{X}$ vs. the alternative that there are heterogeneous treatment effects, i.e., test $H_0 : \boldsymbol{\alpha}_1 = \mathbf{0}$ vs. $H_a : \boldsymbol{\alpha}_1 \neq \mathbf{0}$ in model (13). We use a likelihood ratio test to test this. The test yields a p-value $< 0.001$, providing evidence of heterogeneous treatment effects. When there are heterogeneous treatment effects, the DWH test may not properly control the Type I error rate, but the compliance class likelihood ratio test does properly control the Type I error rate (see Sections 5-6), and hence we will use the compliance class likelihood ratio to test for unmeasured confounding.

Table 8 shows the results of the compliance class likelihood ratio test for unmeasured confounding. There is strong evidence ($p$-value $< 0.001$) that (3) is violated, that is always takers have different risks of death than compliers conditional on the measured confounders $\mathbf{X}$ when both deliver at high level NICUs. There is also evidence that never takers have different risks of death than compliers, that is (4) is violated, but the evidence is not as strong as for always takers (p-value $= 0.012$ compared to p $< 0.001$). In summary, there is strong evidence of some unmeasured confounding.

Table 9 shows, for various combinations of the measured confounders $\mathbf{X}$, the estimated probabilities of death from the model (13) for never takers delivering at low level NICUs, compliers delivering at low level NICUs, compliers delivering at high level NICUs and always takers delivering at high level NICUs. For example, for an infant weighing 1500 grams, whose mother started prenatal care in the second month of pregnancy, whose mother has a high school education and whose mother went into preterm labor, the risk of death is 0.054 for never takers delivering at low level NICUs, 0.051 for compliers delivering at low level NICUs, 0.031 for compliers delivering at high level NICUs and 0.052 for always takers delivering at high level NICUs. This pattern of similar death rates for never takers and compliers delivering at low level NICUs, considerably lower death rates for compliers vs. always takers delivering at high level NICUs and considerably lower death rates for compliers delivering at high level NICUs vs. low level NICUs holds for all combinations of the measured confounders **X**.

|                     | Test Statistic | Degrees of Freedom | p-value   |
|---------------------|:--------------:|:------------------:|:---------:|
| Test of (3)         | 75.9           | 8                  | $< 0.001$ |
| Test of (4)         | 19.6           | 8                  | 0.012     |
| Test of (3) and (4) | 76.6           | 16                 | $< 0.001$ |

**Table 8.** Compliance class likelihood ratio test of no unmeasured confounding for the NICU study.

| Birthweight | Month Prenatal Care Started | Mother's Education | $P_{nt}$ | $P_{co,0}$ | $P_{co,1}$ | $P_{at}$ |
|:-----------:|:---------------------------:|:------------------:|:--------:|:----------:|:----------:|:--------:|
| 1500 | 2 | High School | 0.057 | 0.047 | 0.030 | 0.052 |
| 2000 | 2 | High School | 0.019 | 0.018 | 0.011 | 0.017 |
| 2500 | 2 | High School | 0.006 | 0.006 | 0.004 | 0.005 |
| 1500 | 4 | High School | 0.056 | 0.051 | 0.031 | 0.049 |
| 2000 | 4 | High School | 0.019 | 0.019 | 0.011 | 0.016 |
| 2500 | 4 | High School | 0.006 | 0.007 | 0.004 | 0.005 |
| 1500 | 2 | College     | 0.040 | 0.028 | 0.018 | 0.045 |
| 2000 | 2 | College     | 0.013 | 0.010 | 0.007 | 0.015 |
| 2500 | 2 | College     | 0.004 | 0.004 | 0.002 | 0.005 |
| 1500 | 4 | College     | 0.041 | 0.031 | 0.019 | 0.043 |
| 2000 | 4 | College     | 0.014 | 0.011 | 0.007 | 0.014 |
| 2500 | 4 | College     | 0.004 | 0.004 | 0.002 | 0.004 |

**Table 9.** Estimated Probability of Death For Different Compliance classes with Different Covariate Values. For all sets of covariate values, the mother is assumed to have gone into premature labour, where $P_{nt} = P(Y^0 = 1|C = nt), P_{co,0} = P(Y^0 = 1|C = co), P_{co,1} = P(Y^1 = 1|C = co), P_{at} = P(Y^1 = 1|C = co)$.

## 8. Conclusions and Discussion

We have developed a test of whether there is unmeasured confounding when an instrumental variable (IV) is available. Our test has correct type I error rate unlike the Durbin-Wu-Hausman (DWH) test, which can have inflated type I error rates when there is treatment effect heterogeneity. An important additional advantage of our approach over the DWH test is that it breaks up the test into the two parts (3) and (4), providing more information. For the NICU study, we found evidence that never takers are at a little higher risk of death than compliers when both groups are delivered at low level NICUs and that always takers are at a much higher risk of death than compliers when both groups are delivered at high level NICUs. This latter piece of evidence means that infants who are bypassing local hospitals to go to high level NICUs (i.e., always takers) have unmeasured confounders that makes them have a higher risk of death than infants who would only deliver at a high level NICU if living relatively near to one (i.e., compliers). This suggests that there is some triaging in the way infants are delivering at high level NICUs vs. low level NICUs; future studies could examine how effective this triaging system is.

We have tested the null hypothesis that there is no unmeasured confounding. In some settings, we may instead want to do an equivalence test of the null that the unmeasured confounding is greater than or equal to a specified magnitude vs. the alternative that it is less than this magnitude. For example, instead of testing the null hypothesis of (3), we may want to test

$$H_0 : E(Y^1|C = at, \mathbf{X}) - E(Y^1|C = co, \mathbf{X}) \geq \epsilon \text{ or } \leq -\epsilon \text{ vs.}$$
$$H_a : -\epsilon < E(Y^1|C = at, \mathbf{X}) - E(Y^1|C = co, \mathbf{X}) < \epsilon, \quad (26)$$

where $\epsilon$ is an equivalence margin specified by a subject matter expert. We can test (26) using the two one-sided test procedure [46, 47]. We test $H_0$ in (26) at level $\alpha$ by testing $H_0^1 : E(Y^1|C = at, \mathbf{X}) - E(Y^1|C = co, \mathbf{X}) \geq \epsilon$ and obtaining one-sided p-value $P_1$ and testing $H_0^2 : E(Y^1|C = at, \mathbf{X}) - E(Y^1|C = co, \mathbf{X}) \leq -\epsilon$ and obtaining one-sided p-value $P_2$, and then rejecting $H_0$ if $\max(P_1, P_2) \leq \alpha$. Similarly, we could implement an equivalence test of (4)

The compliance class likelihood ratio test of no unmeasured confounding developed in this paper makes use of assumptions about the probability distribution of the outcome within compliance classes. In this paper, we have considered normal and binary outcomes. In our application, we know that the outcome is binary, but in applications in which the outcome is continuous, we may not know the probability distribution of the outcome. Our test may not perform well when the assumed probability distribution of the outcome does not hold and it is useful to evaluate the goodness of fit of the assumed probability distribution of the outcome. [49] developed an approach to evaluate the goodness of fit for a principal stratification model. If the goodness of fit is not adequate, a different probability distribution model for the outcome can be considered. Our test can easily be extended to non-normal outcome models by using an analogous EM algorithm as in Section 4.1. Also, rather than assuming a parametric model for the outcome distribution, a semiparametric model can be assumed such as the semiparametric density ratio model[7].

In this paper, we have focused on a binary IV. Although mother's excess travel time is a continuous variable, we dichotomized it to be a binary IV (whether the excess time is larger than 10 or not). In practice, it's common that investigators dichotomize multi-level IVs into binary IVs as clinicians may find it easier to think about the validity of the IV assumptions and the interpretation of the IV estimate in terms of a binary IV. However, our method can be extended to a multi-level IV. Suppose the multi-level IV satisfies an extended monotonicity assumption that an individual's potential level of treatment is an increasing function of the level of the IV, $A_i^z \geq A_i^{z'}$ if $z \geq z'$[48]. Let $T_i$ be the smallest $z$ for which $A_i^z = 1$ where $T_i = -\infty$ if $A_i^z = 1$ for all $z$ and $T_i = \infty$ if $A_i^z = 0$ for all $z$. To test for no unmeasured confounding, we can formulate a parametric model for $Y_i^1|T_i, \mathbf{X}_i$ and $Y_i^0|T_i, \mathbf{X}_i$ as in [48], fit the model by maximum likelihood and test (3)-(4).

# 9. Appendix

The complete data log likelihood for the binary outcome is

$$
\begin{aligned}
l_C &= \sum_{i=1}^{n} I(Z_i = 1, A_i = 1) log P(Z_i = 1|X_i) \\
&+ I(Z_i = 1, A_i = 1, C_i = co)[log P(C_i = co|X_i) + log f(Y_i = y|Z_i = 1, C_i = co, X_i)] \\
&+ I(Z_i = 1, A_i = 1, C_i = at)[log P(C_i = at|X_i) + log f(Y_i = y|Z_i = 1, C_i = at, X_i)] \\
&+ I(Z_i = 1, A_i = 0)[log P(Z_i = 1|X_i) + log P(C_i = nt|X_i) + log f(Y_i = y|Z_i = 1, C_i = nt, X_i)] \\
&+ I(Z_i = 0, A_i = 0) log\{1 - P(Z_i = 1|X_i)\} \\
&+ I(Z_i = 0, A_i = 0, C_i = co)[log P(C_i = co|X_i) + log f(Y_i = y|Z_i = 0, C_i = co, X_i)] \\
&+ I(Z_i = 0, A_i = 0, C_i = nt)[[log P(C_i = nt|X_i) + log f(Y_i = y|Z_i = 0, C_i = nt, X_i)] \\
&+ I(Z_i = 0, A_i = 1)[log\{1 - P(Z_i = 1|X_i)\} + log P(C_i = at|X_i) + log f(Y_i = y|Z_i = 0, C_i = at, X_i)].
\end{aligned}
$$

The observed data log likelihood for the binary outcome is

$$
\begin{aligned}
l &= \sum_{i=1}^{n} I(Z_i = 1, A_i = 1) \log P(Z_i = 1|X_i) \\
&+ \sum_{i=1}^{n} I(Z_i = 1, A_i = 1) \log[P(C_i = at|X_i)(P_i^{at})^{y_i}(1 - P_i^{at})^{1-y_i} + P(C_i = co|X_i)(P_i^{C1})^{y_i}(1 - P_i^{C1})^{1-y_i}] \\
&+ \sum_{i=1}^{N} I(Z_i = 1, A_i = 0)[\log P(Z_i = 1|X_i) + \log P(C_i = nt|X_i)(P_i^{nt})^{y_i}(1 - P_i^{nt})^{1-y_i}] \\
&+ \sum_{i=1}^{N} I(Z_i = 0, A_i = 1)[\log P(Z_i = 0|X_i) + \log P(C_i = at|X_i)(P_i^{at})^{y_i}(1 - P_i^{at})^{1-y_i}] \\
&+ \sum_{i=1}^{n} I(Z_i = 0, A_i = 0) \log P(Z_i = 0|X_i) \\
&+ \sum_{i=1}^{n} I(Z_i = 0, A_i = 0) \log[P(C_i = nt|x_i)(P_i^{nt})^{y_i}(1 - P_i^{nt})^{1-y_i} + P(C_i = co|x_i)(P_i^{C0})^{y_i}(1 - P_i^{C0})^{1-y_i}].
\end{aligned}
$$

**Acknowledgements**

This paper is dedicated to the memory of Dr. Thomas R. Ten Have. Dr. Ten Have was a wonderful mentor and role model for Drs. Jing Cheng and Dylan Small. He continued providing encouragement on this research until he passed away. The work of Drs. Jing Cheng and Dylan Small is supported by NIH/NIMH 1RC4MH092722-01 Revised and Jing Cheng was also supported by grant NIH/NIDCR U54DE019285.

# References

1. Rosenbaum, P.R. *Observational Studies*, 2001, 2nd ed. Springer, New York.

2. Angrist, J.D., Imbens, G.W. and Rubin, D.R. Identification of causal effects using instrumental variables (with discussion). *Journal of the American Statistical Association* 1996; **91**, 444-472.
3. Abadie, A. Bootstrap tests for distributional treatment effects in instrumental variable models. *Journal of the American Statistics Association* 2002; **97**, 284-292.
4. Hernán, M.A. and Robins, J.M. Instruments for causal inference: an epidemiologist's dream?. *Epidemiology* 2006; **14**, 360-372.
5. Tan, Z. Regression and weighting methods for causal inference using instrumental variables. *Journal of the American Statistical Association* 2006; **101**, 1607-1618.
6. Brookhart, M.A. and Schneeweiss, S. Preference-based instrumental variable methods for the estimation of treatment effects: assessing validity and interpreting results. *International Journal of Biostatistics* 2007; **3**.
7. Cheng, J., Qin, J. and Zhang, B. Semiparametric estimation and inference for distributional and general treatment effects. *Journal of the Royal Statistical Society: Series B* 2009;**71**, 881-904.
8. Cheng, J., Small, D.S., Tan, Z. and Ten Have, T.R. Efficient nonparametric estimation of causal effects in randomized trials with noncompliance. *Biometrika* 2009; **96**, 1-9.
9. Baiocchi, M., Small, D.S., Lorch, S. and Rosenbaum, P.R. Building a stronger instrument in an observational study of perinatal care for premature infants. *Journal of the American Statistical Association* 2010; **105**, 1285-1296.
10. Durbin, J. Errors in variables. *Review of the International Statistical Institute* 1954; **22**, 23-32.
11. Wu, D.-M. Alternative tests of independence between stochastic regressors and disturbances. *Econometrica* 1973; **41**, 733-750.
12. Hausman, J. Specification tests in econometrics. *Econometrica* 1978;**41**, 1251-1271.
13. Wooldridge, J.M. *Econometric Analysis of Cross Section and Panel Data*, 2010. 2nd Edition. Cambridge: MIT Press.
14. Basu, A., Heckman, J.J., Navarro-Lozano, S. and Urzua, S. Use of instrumental variables in the presence of heterogeneity and self-section: an application to treatments of breast cancer patients. *Health Economics* 2007; **16**, 1133-1157.
15. Brookhart, M.A., Rassen, J.A. and Schneeweiss, S. Instrumental variable methods in comparative safety and effectiveness research. *Pharmacoepidemiology and Drug Safety* 2010;**19**, 537-554.
16. Tan, Z. Marginal and nested structural models using instrumental variables. *Journal of the American Statistical Association* 2010; **105**: 157-169.
17. Hahn, J., Ham, J.C. and Moon, H.R. The Hausman test and weak instruments. *Journal of Econometrics* 2011; **160**, 289-299.
18. Adkins, L.C., Campbell, R.C., Chmelarova, V. and Hill, R.C. The Hausman test, and some alternatives with heteroskedastic data. In *Essays in Honor of Jerry Hausman: Advances in Econometrics, Volume 29*, eds., Baltagi, B.H., Newey, W.K. and White, H.L. Emerald Books, Warwick, United Kingdom.
19. Lorch, S., Myers, S. and Carr, B. The regionalization of pediatric health care. *Pediatrics* 2010; **126**, 1182-1190.
20. Howell, E.M., Richardson, D., Ginsburg, P. and Foot, B. Deregionalization of neonatal intensive care in urban areas. *American Journal of Public Health* 2002; **92**, 119-124.
21. Lorch, S.A., Baiocchi, M., Fager, C. and Small, D. The differential impact of delivery hospital on the outcomes of premature infants. *Pediatrics*, in press and published online, doi: 10.1542/peds.2011-2820 .
22. McClellan, M., McNeil, B., Newhouse J. Does more intensive treatment of acute myocardial infarction in the elderly reduce mortality? analysis using instrumental variables. *Journal of the American Medical Association* 1994;**272**, 859-866.
23. Phibbs, C.S., Mark, D.H., Luft, H.S., Peltzmanrennie, D.J.,Garnick, D.W. et al. Choice of hospital for delivery – a comparison of high-risk and low-risk women.*Health Services Research* 1993;**28**, 201-222.
24. Frangakis, C.E. and Rubin, D.B. Addressing complications of intention-to-treat analysis in the combined presence of all-or-none treatment-noncompliance and subsequent outcomes. *Biometrika* 1999;**86**, 365-379.
25. O'Malley, A.J. and Normand, S.-L.T. Likelihood methods for treatment noncompliance and subsequent nonresponse in randomized trials. *Biometrics* 2005; **61**, 325-334.
26. Zhou, X.-H. and Li, S.M. ITT analysis of randomized encouragement design studies with missing data. *Statistics in Medicine* 2006; **25**, 27372761.
27. Jo, B., Asparouhov, T. and Muthén, B.O. Intention-to-treat analysis in cluster randomized trials with noncompliance. *Statistics in Medicine* 2008; **27**, 5565-5577.
28. VanderWeele, T.J. and Arah, O.A. Bias formulas for sensitivity analysis of unmeasured confounding for general outcomes, treatments, and confounders. *Epidemiology*, 2011, **22**, 42-52.
29. Rubin, D.B. Statistics and causal inference: comment: which ifs have causal answers. *Journal of the American Statistical Association* 1986;**81**, 961-962.
30. Cole S R, Frangakis C E. The consistency statement in causal inference: a definition or an assumption?. *Epidemiology*, 2009, **20(1)**: 3-5.
31. Rosenbaum, P.R. and Rubin, D.B. The central role of the propensity score in observational studies for causal effects. *Biometrika* 1983;**70**: 41-55.

32. Imbens, G. Nonparametric estimation of average treatment effects under exogeneity: A review. *Review of Economics and Statistics* 2004;**86**, 1-29.
33. Joffe, M. and Brensinger, C. Weighting in instrumental variables and g-estimation. *Statistics in Medicine* 2003: **22**, 1285-1303.
34. Small, D., Ten Have, T., Joffe, M. and Cheng, J. Random effects logistic models for analysing efficacy of a longitudinal randomized treatment with non-adherence. *Statistics in Medicine*, 2006, **25**, 1981-2007.
35. Imbens, G. W. and Rubin, D. B. Estimating outcome distributions for compliers in instrumental variables models. *Review of Economic Studies* 1997;**64**, 555-574.
36. Imbens, G. W. and Rubin, D. B. Bayesian inference for causal effects in randomized experiments with noncompliance. *The Annals of Statistics* 1997; **25**, 305-327.
37. Casella, G. and Berger, R.L. *Statistical Inference*, 2001, second edition, page 375.
38. Angrist, J.D. and Krueger, A.B. Empirical Strategies in Labor Economics, in *Handbook of Labor Economics*, 1999, O. Ashenfelter and D. Card eds., Elsevier: Amsterdam.
39. Lange, K. A quasi-Newton acceleration of the EM algorithm. *Statistica Sinica* 1995; **5**, 1-18.
40. Jamshidian, M. and Jennrich, R.I. Acceleration of the EM algorithm by using quasi-Newton methods. *Journal of the Royal Statistical Society, Series B* 1997; **59**, 569-587.
41. Press, W.H., Teukolsky, S.A., Vetterling, W.T. and Flannery, B.P. *Numerical Recipes. The Art of Scientific Computing*, 2007. Third edition. Cambridge University Press: New York.
42. R Core Team, *R: A Language and Environment for Statistical Computing*, R Foundation for Statistical Computing, 2013. http://www.R-project.org/.
43. Bowden, R.J. and Turkington, D.A. *Instrumental Variables*, 1984, Cambridge University Press: New York.
44. White, H. *Asymptotic Theory for Econometricians*, 1984, Academic Press: New York.
45. VanderWeele, T.J. and Hernan, M.A. Causal inference under multiple versions of treatment. *Journal of Causal Inference* 2013; **1**, 1-20.
46. Schuirmann, D. L. On Hypothesis Testing to Determine if the Mean of a Normal Distribution Is Contained in a Known Interval. *Biometrics*, 1981, **37**, 617.
47. Westlake, W.J. Response to T.B.L. Kirkwood: bioequivalence testing - a need to rethink. *Biometrics*,1981, **37**, 589-594.
48. Glickman, M. E. and Normand, S.T. The derivation of a latent threshold instrumental variables model. *Statistica Sinica*,2000,**10**,2,517-544.
49. Zhang, J. L., Rubin, D. B. and Mealli, F. Likelihood-based analysis of causal effects of job-training programs using principal stratification. *Journal of the American Statistical Association*,2009,**104**,485,166-176.